Correspondence and requests for materials. Amelia Barreiro ab3690@columbia.edu; Felix Börrnert fb1@ifw-dresden.de

# Understanding the catalyst-free transformation of amorphous carbon into graphene by current-induced annealing

Amelia Barreiro*,[†,1], Felix Börrnert*,[†,2], Stanislav M. Avdoshenko[3,4], Bernd Rellinghaus[2], Gianaurelio Cuniberti[3], Mark H. Rümmeli[2,3], Lieven M. K. Vandersypen[1]

[†] both authors contributed equally
[1] Kavli Institute of Nanoscience, Delft University of Technology, Lorentzweg 1, 2628 CJ Delft, The Netherlands
[2] IFW Dresden, Postfach 270116, 01171 Dresden, Germany
[3] TU Dresden, 01062 Dresden, Germany
[4] School of Materials Engineering, Purdue University, West Lafayette, IN, USA

**Abstract**

We shed light on the catalyst-free growth of graphene from amorphous carbon (a-C) by current-induced annealing by witnessing the mechanism both with in-situ transmission electron microscopy and with molecular dynamics simulations. Both in experiment and in simulation, we observe that small a-C clusters on top of a graphene substrate rearrange and crystallize into graphene patches. The process is aided by the high temperatures involved and by the van der Waals interactions with the substrate. Furthermore, in the presence of a-C, graphene can grow from the borders of holes and form a seamless graphene sheet, a novel finding that has not been reported before and that is reproduced by the simulations as well. These findings open up new avenues for bottom-up engineering of graphene-based devices.

Introduction

Graphene, a single atomic layer of carbon connected by sp$^2$ hybridized bonds, has attracted intense scientific interest since its recent discovery [1]. Much of the research on graphene has been directed towards the exploration of its novel electronic properties which open up new avenues to both exciting experiments in basic science [2-5], and electronic applications [6]. Further experiments and novel devices could be envisaged but remain to be demonstrated due to technological challenges in fabrication such as the lack of precision for locating or growing graphene of a specific size on a substrate of choice.

Whilst significant strides have been made in understanding graphene synthesis [7], the mechanisms behind growth remain highly debated. Graphene growth cannot be captured by a universal mechanism with specific routes and conditions but a variety of synthesis strategies and growth modes exist. The best-known mechanism is the use of metal catalysts whereby free carbon radicals are formed, carbon is dissolved in the catalyst, and finally precipitates at the surface. The free carbon radicals usually are supplied from a hydrocarbon feedstock, but there also are a few reports where the carbon feedstock is provided by a-C [5-10].

Another surface that can provide suitable sites for growth is a bulk oxide support without any metal catalyst present where the carbon precursor is supplied by a hydrocarbon feedstock [11-14]. In the case of graphene growth from stable oxides as the support material, carbon dissolution is unlikely and therefore the growth probably occurs through surface diffusion processes. Oxides without a metal catalyst can also be used for the growth of carbon nanotubes (CNTs) [15-16].

The growth of sp$^2$ structures without a catalyst relies on a mechanism that largely remains to be understood [17]. Another example of such a process is the formation of CNTs on the cathode in the arc-discharge route which can occur without catalyst addition above 4000 ºC [18-21]. More recently, other growth routes without catalyst have emerged such as the formation of CNTs on graphitic surfaces [22,23], the substrate-free gas-phase synthesis of graphene sheets [24], or the growth of graphene sheets by microwave chemical vapour deposition (CVD) [25].

Recently, the non-catalytic graphitization of a-C into small (~ 10 nm) polycrystalline graphene [26], and into additional shells on multi-walled (MW) CNTs [21,27,28] by current-induced annealing of graphene or of MWCNTs, respectively, has been reported. Moreover, catalyst-free crystallization of a-C nanowires led to the formation of tubular graphitic shells with nano-onions in their interior [29]. Unfortunately, the quality of all these graphitized nanostructures was rather poor as compared to arc-discharge grown CNTs or mechanically exfoliated graphene, presumably because temperatures were insufficiently high (below 3000 ºC) to induce perfect graphitization [21,30]. Additionally, a recent theoretical report points towards template assisted graphene growth [31].

In this Article we report on in-situ transmission electron microscopy (TEM) studies of the structural changes that lead from a-C to crystalline graphene patches of over 10 x 10 nm in size, and to even larger patches up to 100 x 300 nm. Furthermore, we use molecular dynamics (MD) simulations in order to get more insight in the process that transforms a-C to graphene when on top of a graphene substrate. Both in experiment and in theory, we observe that small a-C clusters on top of a graphene substrate rearrange and crystallize into graphene patches. The process is aided by the high temperatures involved and by the van der Waals [32] interactions with the substrate. Finally, in the presence of a-C, graphene can grow from the borders of holes and form a seamless graphene sheet, a novel finding that has not been reported before and that is reproduced by the simulations as well.

Results section

We perform in-situ current-induced annealing of suspended graphene devices by taking the samples to the high bias regime, specifically up to 2 - 3 V by stepwise increasing the voltage bias in 10 mV steps [33, 34], please see the Methods Section for further details. In this regime, the samples are at such a high bias and, therefore such a high current is flowing through them, that they are close to a complete and irreversible electrical breakdown and we start to sublime different regions and layers of the graphene device [35]. *Via* this procedure contaminants from the fabrication process are also effectively removed as they were on top of the removed layers [35], and we observe that we obtain atomically clean graphene devices, as can be resolved from TEM imaging. After the current annealing process, the bias is taken back to 0 V and the samples cool down. Exposure of the cold samples to the electron beam allows us to intentionally deposit a-C on the previously clean graphene surfaces [36], see figure 1 and section S1 in the Supporting Information. The carbon source originates from the beam-aided decomposition of hydrocarbons in the TEM column and/or from organic impurities adsorbed on the chip, the chip carrier and the sample holder. The regions where the a-C preferentially deposits are the edges of the individual layers in few layer graphene flakes, edges and other defects, fig. 1 [37]. Amorphisation of the graphene sheet because of disorder introduced by the electron beam is unlikely at an acceleration voltage of 80 keV, which is below the "knock-on" damage threshold of carbon nanostructures, see section 3 in the Supporting Information [38, 39]. Thus graphene sheets remain stable and defect free in clean regions [40]. However, holes can form in contaminated areas by beam-driven chemical modifications with contaminants and adsorbates at energies below the knock-on threshold [41]. These holes seem to concentrate around edges and other defects, fig. 1 b, c [37]. Interestingly, during the current-annealing process itself, we never observe deposition of a-C. Presumably, hydrocarbon precursors for a-C formation desorb before being able to reach the graphene flake due to the high temperatures and deposit on colder areas around the hot graphene.

After deposition of a-C on the previously atomically clean graphene surfaces, the samples are brought back once more to the high bias regime, specifically up to 2 - 3 V, and current-annealed again without reaching the high-current limit where also the graphene substrate starts to sublime [35]. We proceed by stepwise increasing the voltage, wait for changes to occur and then slowly increase the bias voltage further. Temperatures as high as 2000 ˚C [26, 42], or even 3000 ˚C [29], have been estimated to be reached due to Joule heating. Surprisingly, during this process we observe that it is not possible to sublime the a-C but instead it gradually transforms into graphene patches. We have observed the transformation of a-C to graphene by current-induced annealing on 15 of 15 samples where a-C had been intentionally deposited by the TEM beam. Figures 2 and 3 illustrate the evolution of the process from amorphous matter to crystalline few-layer graphene terraces.

Due to the nature of our in-situ TEM experiments, we can unequivocally testify to the circumstances during growth by performing atomic resolution imaging. Small a-C clusters rearrange and crystallize due to the high temperatures reached during current annealing without the involvement of any catalyst. Before reaching temperatures high enough to sublime the a-C, it gradually rearranges into high-quality graphene, see figures 3 and 4. The supplementary information (SI) contains low magnification TEM images (section S-4) and a video (movie S-1) of a different device where an overview can be obtained regarding the gradual transformation of a-C to graphene by current-induced annealing.

Based on high resolution (HR) TEM (see Figure 4) we were able to confirm that indeed the newly grown patches are graphene. From the corresponding Fourier transform (FT) in Fig. 4 (b) we can obtain the typical lattice parameter of graphene and the orientation of the newly grown layer which is rotated by 22 degrees with respect to the substrate. These patches can reach more than 100 nm x 100 nm in size (see figure S-4). The fact that we obtain a clear FT signal from an overlayer of an area of approx. 30 nm$^2$ suggests that the graphene is not disordered and has a "long range order", i.e. consists of a single grain.

Another interesting finding is that in the presence of a-C at high bias it is possible to repair holes in the graphene lattice. In Figure 5 we observe that holes formed by the reaction of contaminants with graphene due to the electron beam [41], are self-repaired by growing new graphene healing the holes. Recently, it was found that multi-vacancies in a graphene lattice can be quickly reoccupied by C ad-atoms and graphene can recover its crystallinity. This repairing mechanism works best at temperatures above 600 °C and was attributed to lattice reconstructions [36] but can also occur spontaneously [43]. Healing of multivacancies in carbon nanotubes with up to 20 missing atoms can also be achieved with lattice reconstructions due to the TEM beam [44]. The holes in our graphene lattice are much bigger and can have diameters up to 5 nm [37]. Indeed, figure 5 suggests that the holes are closing step by step, presumably by the formation of new bonds with carbon radicals originating from the a-C. In this sense, the healing mechanism of the holes can be understood as substrate-free growth starting from the borders of holes in the graphene lattice by carbon atom addition to the reactive dangling bonds at the edges.

Low magnification TEM images (figure S-4 in the electronic supplementary material) and a video (movie S-2) of a different device showing an overview regarding the gradual healing of holes in graphene in the presence of a-C by current-induced annealing can be found in the SI [39].

Discussion section

In order to shed light on the catalyst-free transformation mechanism of a-C into graphene and to explore the role of the graphene substrate, we performed molecular dynamics simulations of a perfect graphene substrate and four a-C clusters of 1 nm diameter on top at a distance of ~ 3.5 Å (Figure 6); for more details see the supporting information. During the MD simulations, the graphene substrate and the four a-C clusters on top are subject to stepwise increasing temperatures, namely, 300, 600, 1200 and 1800 K. This scenario resembles our experimental procedure well. In the experiments, a temperature distribution with a maximum close to the center of the graphene flake is present because the heat is only evacuated through the electrodes.

In our theoretical model we assume a constant temperature, which is a reasonable approximation for the small windows used in the simulations, since within the hot spot the thermal gradient is small. Upon increasing temperature, the a-C starts transforming, goes through a glass-like phase in the range 600 - 1200 K and finally forms a graphene structure at 1800 K, see figure 6 b [37]. Indeed, the need for higher temperatures to overcome Stone-Wales barriers to obtained perfect hexagonal graphene lattices has recently been reported [31]. The structure formed is flat and is located 3.5 Å above the initial graphene template. Nevertheless, holes are still present because insufficient carbon feed material was available to grow graphene over the whole area of the underlying graphene template. Upon further addition of a-C at 1800 K the graphene structure grows and, defects are progressively healed out (Fig. 6c). Indeed, some areas display defect-free graphene such as in figure 6d. The remaining holes and defects could be healed by further addition of a-C and a longer annealing time but this would require excessively long calculation times. In movie S-3, which shows the transformation of a-C to graphene, one can observe that only in the final stages of the growth process when the graphene precursor flakes merge into a bigger unit, they become stationary on the surface. This is due to the energy gain provided by the π-π coupling, suggesting that any atomically smooth substrate could serve as a template.

Recently, similar MD simulations modeled the synthesis of fullerenes [45-47]. An important difference between those simulations and ours is that in our case there is a graphene substrate while the fullerene synthesis was obtained in a substrate free model. Regardless of the initial geometry and velocity of the a-C, at the end of the MD runs, graphene is reproducibly formed on top of the graphene template. Although the substrate is only weakly coupled to the a-C, it apparently strongly influences the transformation of the nanostructure on top of it and prevents the formation of fullerene-like structures, demonstrating its influence on the formation of graphene. Experimental evidence that confirms the graphene template is only weakly coupled to a-C is given by the fact that the newly grown graphene patch is rotated with respect to the initial support layer, see figure 4. Again, these results suggest that in a more general picture our graphene growth method is universal for atomically smooth template-supported processes such as graphene, hBN or other two-dimensional layered materials such as $MoS_2$. Indeed, a similar "substrate effect" has recently been recently reported for the growth of graphene on Ni [31]. In this theoretical study, the authors have predicted that Haeckelite is preferentially nucleated from ensembles of $C_2$ molecules on a clean Ni(111) face, with graphene as a metastable intermediate phase. To the contrary, in the presence of a coronene-like $C_{24}$ template, hexagonal ring formation is clearly promoted and finally anneal to graphene. Experimentally, in another study it was possible to grow graphene on hBN by CVD [48], further supporting the universality of this growth method on atomically flat 2D systems.

To gain further insight into the experimental observation of healing holes, we performed MD simulations. We create a hole with a 1 nm radius in an ideal graphene flake and place 3 a-C clusters (of 1 $nm^3$ size each) on top of it, see figure 7. We then heat our system to 1800 K. First, long fibers and big polyedres ($C_{8-10}$) are formed across the hole. With further annealing, the hole is healed completely. For several independent runs with different initial structural and velocity conditions it takes 25 - 30 ps to completely heal the hole. The newly grown graphene contains at least one Stone-Wales defect (two pentagons ($C_5$) and two heptagons ($C_7$) forming a double pair)

[37], which we anticipate would fully heal out if significantly longer simulation time were available. This process of graphene forming in and healing a hole can be seen in movie S-4 [39].

Interestingly, running the same MD simulations at temperatures below 600 K instead of at 1800 K does not yield healed out holes, suggesting that higher temperatures are required for repairing holes. Effective changes in a reasonable timeframe for the simulations (around 10 ps) only take place above 600 K. Due to the high temperatures reached during current annealing, untangled and unsaturated bonds from the a-C diffuse on the graphene and act as a source of radicals. They react with the dangling bonds at the edge of holes, gradually healing them out forming a new graphene lattice.

The speed of the transformation from a-C to graphene or the growth of graphene in holes in our MD simulations can be markedly fast, down to about 50 ps. However, the time elapsed for the transformation of a-C to graphene observed experimentally takes up to 1 - 15 minutes. The large difference in velocity between the experiments and simulations is attributed to the difference in the system dimensions, note that the graphene patches in the simulations are approx. 5x5 nm$^2$ in size and still contain many holes and defects, which would heal out if the simulation times could be significantly increased. In contrast, the patches grown experimentally can be as big as 100 x 300 nm$^2$, see figure S5 in the Supporting Information [39]. Moreover, the experiments were conducted as slowly as possible to prevent bringing the sample to an excessively high bias where the transformation would occur quicker but the risk of a complete electrical breakdown of the sample is larger.

In conclusion, our *in situ* real-time TEM observations correlated with MD simulations shed light on the catalyst-free transformation of a-C to flat graphene sheets. Small a-C clusters rearrange and crystallize into graphene at high temperatures on a graphene substrate or from the edges of holes. This finding opens up new avenues for engineering novel graphene-based devices in which additional graphene layers are needed on top of a graphene substrate [49, 50]. To this end, clusters of a-C could be deposited on specified locations on top of graphene *via* e-beam deposition [51], and then be transformed to additional graphene patches *in-situ* by a further (current-

)annealing step. In a more general picture, our graphene growth method seems to be universal for atomically smooth template-supported processes such as graphene, hBN or other strongly layered 2D crystals such as $MoS_2$ or $NbSe_2$. Transforming a-C to graphene could open up new avenues for novel devices consisting of graphene on top of 2D materials of choice.

Methods section

Chips with single-layer and few-layer graphene flakes supported by metal contacts were mounted on a custom-built sample holder for TEM with electric terminals, enabling simultaneous TEM imaging and electrical measurements. For imaging, a FEI Titan$^3$ 80–300 transmission electron microscope with a CEOS third-order spherical aberration corrector for the objective lens was used. It was operated at an acceleration voltage of 80 kV to reduce knock-on damage. All studies were conducted at room temperature with a pressure of approx. $10^{-7}$ mbar. The graphene device fabrication and measurement procedures are described in detail in references 32 and 53. In brief, a graphene flake is transferred onto Cr/Au electrodes that are freely suspended over an opening in a $Si/SiO_2$ wafer. The device is voltage biased and the current is measured. In total, we measured 15 devices, with spacings between the electrodes between 1 and 20 μm.

Acknowledgments

We gratefully acknowledge M. Rudneva and H. Zandbergen for help in the early stages of the experiment, G. F. Schneider for help with graphene transfer and M. Zuiddam for help with the deep reactive ion etching process. Financial support was obtained from the Dutch Foundation for Fundamental Research on Matter (FOM), Agència de Gestió d´Ajuts Universitaris i de Recerca de la Generalitat de Catalunya (2010_BP_A_00301), DFG (RU1540/8-1), EU (ECEMP) and the Freistaat Sachsen.

Author contribution statement

A. B. fabricated the samples and performed the electronic measurements. F. B. performed the TEM measurements. S. A. M performed the molecular dynamis simulations. A. B. wrote the manuscript. All authors discussed and commented on the manuscript.

Additional information. Supplementary Information available. The authors declare no competing financial interests.

Figure caption

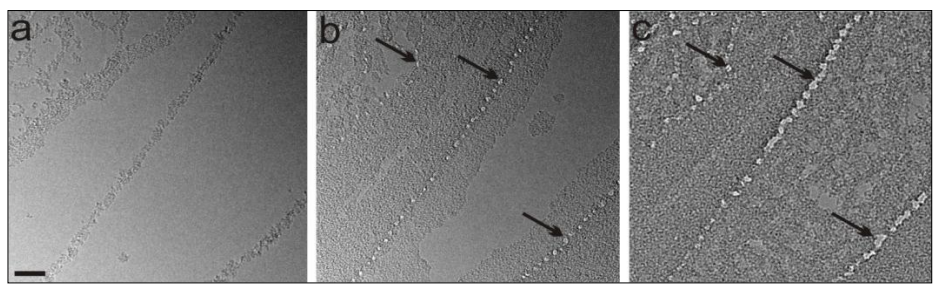

**Figure 1**. TEM images of the stepwise deposition of a-C on an initially clean graphene sheet due to imaging. The scale bar is 20 nm. (a) Preferential deposition of a-C at edges of other graphene layers. The a-C is the darker and rough surface. (b) Formation of holes (bright spots in the images, marked with arrows) and further deposition of a-C. (c) Growth of holes (marked with arrows) and almost complete coverage of a-C on the graphene template.

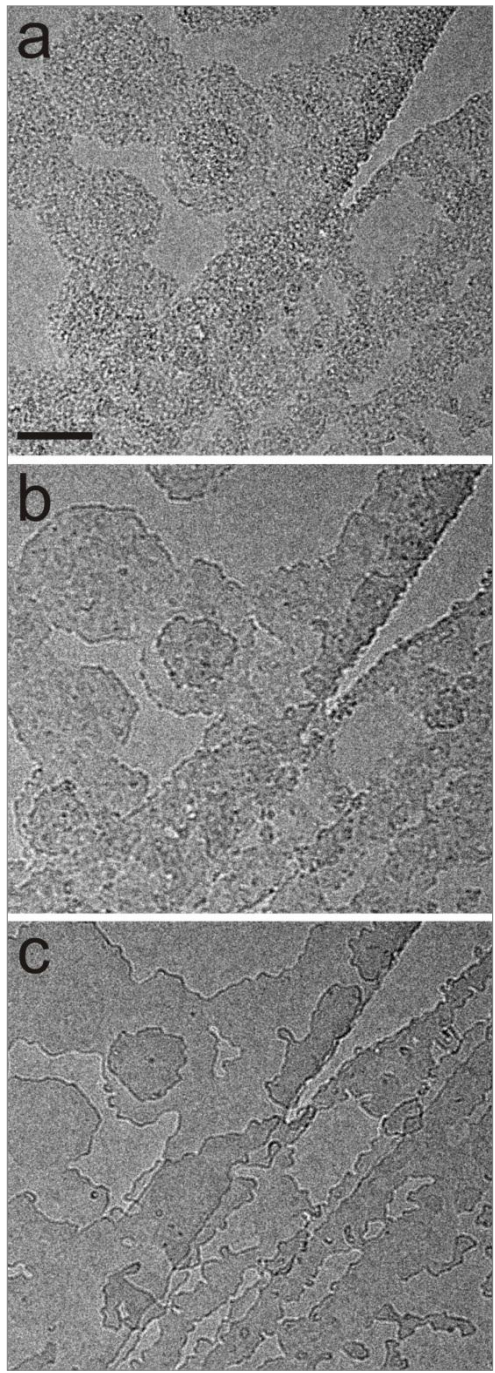

**Figure 2.** TEM images of the stepwise transformation of a-C into few-layer graphene terraces by means of current annealing. (a) a-C on graphene. The scale bar is 20 nm. (b) Gradual crystallization of the a-C through a glass-like phase (2.2 V, 0.4 mA/µm). (c) Transformation into graphene patches (2.68 V, 0.8 mA/ µm). The time elapsed between frame (a) to (c) is 5 min and 30 seconds.

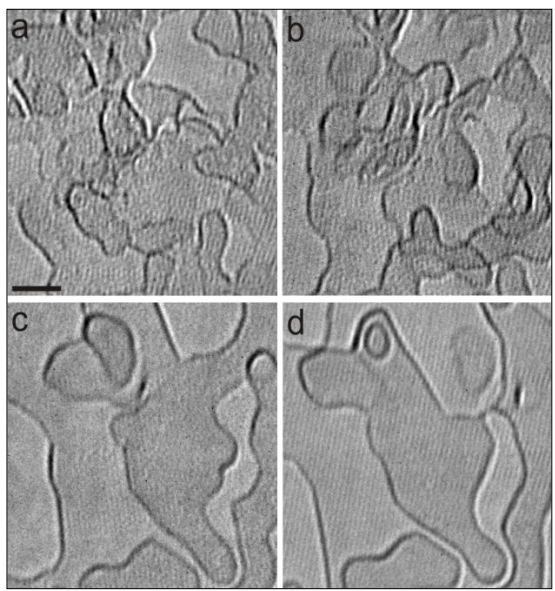

**Figure 3.** Aberration corrected HR-TEM images of the stepwise transformation of a-C into graphene patches by means of current annealing at 3.32 V, 0.55 mA. (a) a-C on graphene. The scale bar is 2 nm. (b,c) Gradual crystallization of the a-C through a glass-like phase, see figure S6 in the Supplementary Information for further illustration of this process [39]. (d) Transformation into highly ordered graphene patches. The time elapsed between the four frames is 22 minutes.

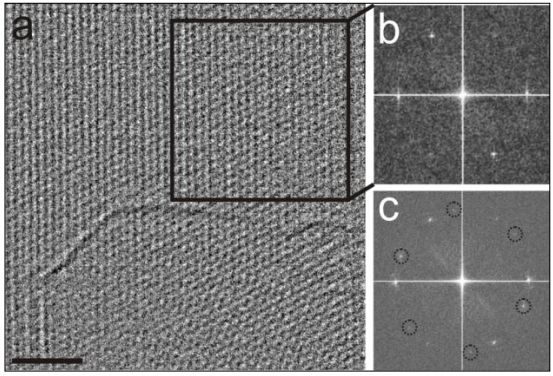

**Figure 4.** HR-TEM images of the transformation of a-C to graphene and the graphene supporting layer. (a) Aberration corrected HR-TEM image of the graphene support (top) and a graphene patch grown on top of it (bottom) from a-C by means of current annealing. The scale bar is 2 nm. The contrast of the micrograph was enhanced through Wiener filtering to suppress noise. Note that obtaining high resolution TEM images at such high bias voltages is challenging due to thermal vibrations. (b) FT of the initial graphene layer in the upper part of the TEM image marked by a square. (c) FT of the whole micrograph containing the graphene support layer and the graphene patch grown on top (circles), which are rotated with respect to each other by 22º.

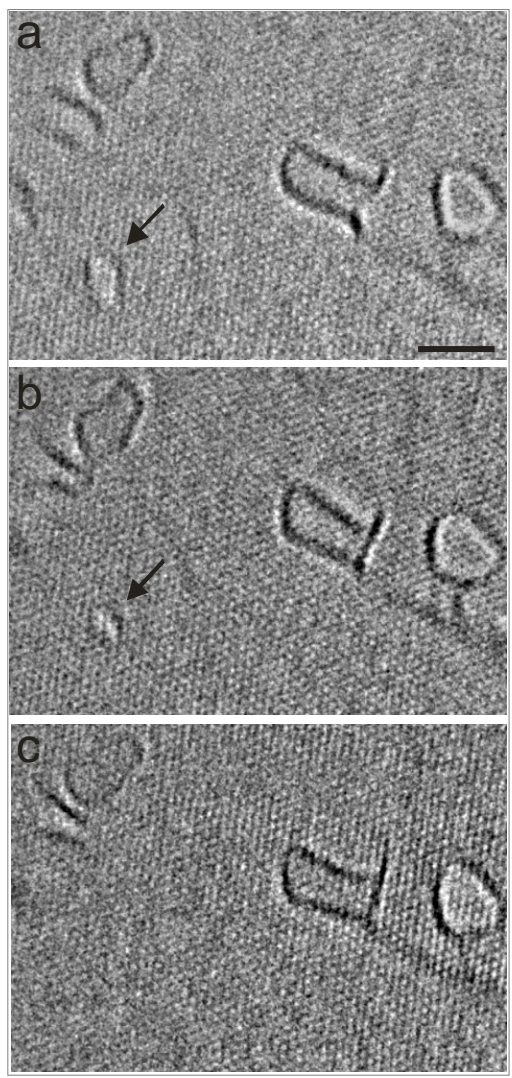

**Figure 5.** Aberration corrected HR-TEM images of the gradual healing of a hole in the graphene lattice by means of current annealing of graphene in the presence of a-C at 2.75 V, 2.2 mA. The arrows point to the initial hole (a), that gradually gets smaller (b) until it completely heals out (c). The scale bar is 2 nm. The time elapsed between the three frames is 30 s.

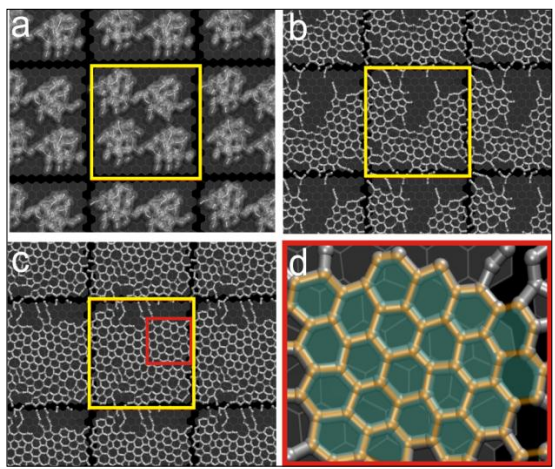

**Figure 6.** Molecular dynamics simulations of the stepwise transformation of a-C to graphene. (a) Initial 4 a-C clusters on top of a graphene unit cell marked by a yellow square. (b) Intermediate stage after annealing at 1800 K. (c) Structural shape after further a-C addition at 1800 K. The time elapsed between each frame is ~50 ps (see text for details). Further annealing at 1800 K with more carbon feedstock helped to overcome pentagon/heptagon defect structures, as high temperatures help to overcome Stone-Wales barriers to obtain hexagonal graphene lattices [31]. (d) Zoom in into the red square in panel (c) displaying a perfect graphene region without defects.

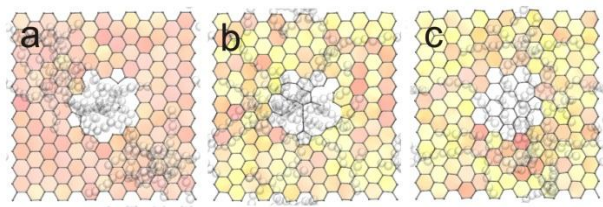

**Figure 7.** Molecular dynamics simulations at 1800 K describing the hole healing process. a-C is represented by the semi-transparent spheres. (a) Initial configuration displaying a hole in the graphene lattice. (b) Intermediate state of the structural transformation. (c) Repaired hole. The colors represent the actual number of vertices of the rings and their bending order. The full color code is described in Table 1, page 134 in Ref. 52.

# Supplementary Information

# Understanding the catalyst-free transformation of amorphous carbon into graphene by current-induced annealing

Amelia Barreiro*,[†,1], Felix Börrnert*,[†,2], Stanislav M. Avdoshenko[3,4], Bernd Rellinghaus[2], Gianaurelio Cuniberti[3], Mark H. Rümmeli[2,3], Lieven M. K. Vandersypen[1]

[†] both authors contributed equally
[1] Kavli Institute of Nanoscience, Delft University of Technology, Lorentzweg 1, 2628 CJ Delft, The Netherlands
[2] IFW Dresden, Postfach 270116, 01171 Dresden, Germany
[3] TU Dresden, 01062 Dresden, Germany
[4] School of Materials Engineering, Purdue University, West Lafayette, IN, USA

## Section 1. Experimental methods

We have performed in-situ transmission electron microscopy (TEM) studies of graphene growth from a-C by current-induced annealing. In order to observe the dynamics of these structural changes the chip with the contacted graphene sample was mounted on a custom-built sample holder for TEM with electric terminals, enabling simultaneous TEM imaging and electrical measurements. For imaging, a FEI Titan[3] 80–300 TEM with a CEOS third-order spherical aberration corrector for the objective lens was used. It operated at an acceleration voltage of 80 kV to reduce knock-on damage. The images were recorded with a Gatan UltraScan 1000 camera via the Gatan DigitalMicrograph software. To enhance the temporal resolution for in situ observation down to 350 ms per frame the camera was used in conjunction with the TechSmith Camtasia Studio screen recorder software at 4 pixel binning with an acquisition time of 0.05 s.

**Section 2. Preferential deposition of a-C on the edges of individual layers and defects in few layer graphene**

The source of a-C can originate from the decomposition of hydrocarbons in the TEM column and/or from hydrocarbon-containing organic impurities adsorbed on the chip, the chip carrier and the sample holder. The mobile hydrocarbons diffuse and reach the area exposed to the electron beam. Under electron irradiation at 80 kV, mobile hydrocarbon deposits are converted to a-C, while hydrogen atoms are knocked out by electron impacts. We observed on 3 samples that preferential sites for a-C adsorbtion and consequent a-C formation are edges or defect sites of few layer graphene, see "Fig. S1". A possible explanation for this finding is that during the diffusion of the mobile hydrocarbons on the graphene surface, they encounter an obstacle when reaching an edge or a defect and tend to adsorb there. As a result, a-C deposits form and gradually grow bigger.

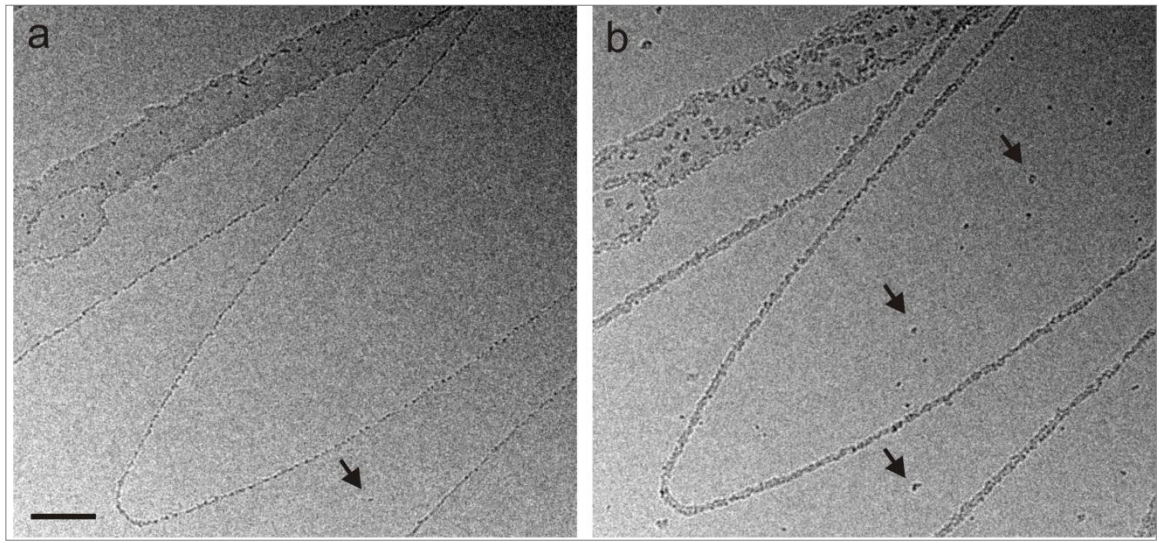

**Figure S1.** TEM images of the stepwise deposition of a-C. (a,b) At the edges of the individual layers in few layer graphene flakes and (b) other defects. The arrows point at defects in the lattice where a-C agglomerates while the continuous lines with elliptical shapes separate regions with a different number of layers. a-C preferentially deposits at the step edges. The scale bar is 50 nm.

**Section 3. Effect of the 80 keV electron beam**

In all the experiments we image with an 80 keV beam. It has been reported that an 80 keV beam can lead to the healing of multivacancies with up to 20 missing atoms by lattice reconstructions [1]. Nevertheless, for free-standing graphene, the electron beam can only heat up the sample in the order of about 1 K [2]. In contrast, current annealing in the high current limit was estimated to lead to temperatures as high as 2000 - 3000 °C [3-5]. Therefore, the main driving mechanism to the formation of graphene from a-C should originate from Joule heating, although we cannot exclude that there might be a contribution to the transformation from the electron beam. Indeed, the electron beam does induce crystallization of the amorphous carbon [6], but the time scale of the conversion is about one order of magnitude slower.

Importantly, if TEM imaging was performed without concomitant current annealing we never observed the formation of graphene out of a-C but only the deposition of increasing quantities of a-C. Moreover, if as-fabricated graphene flakes are imaged without a previous current annealing step, residues from fabrication react with the graphene by beam-driven chemical modifications with contaminants and adsorbates at energies below the knock-on threshold [7] leading to the rupture of the flakes, see "Fig. S2".

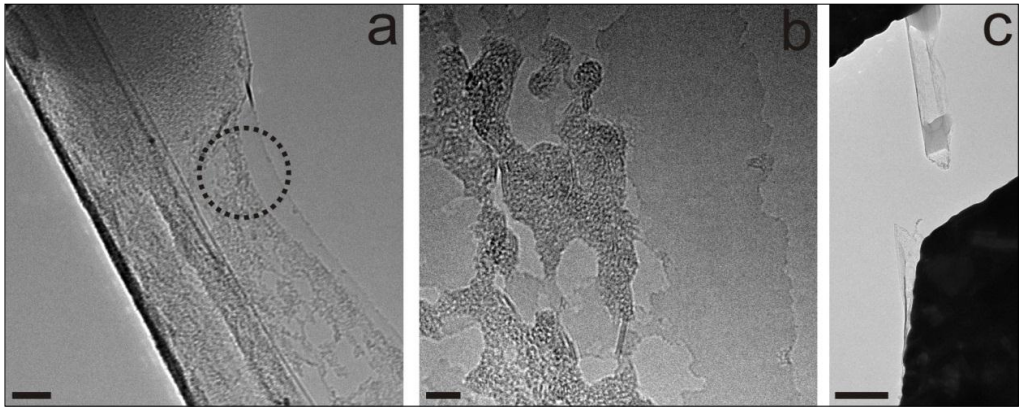

**Figure S2**. Effect of the 80 keV beam. (a) TEM image of a few-layer graphene flake contaminated by fabrication residues. The scale bar is 20 nm. (b) Hole formation due to beam-driven chemical modifications in the lattice with the contaminants. The scale bar is 5 nm. (c) Rupture of the graphene due to an excessive accumulation of holes. The scale bar is 200 nm.

**Section 4. Transformation of a-C to graphene**

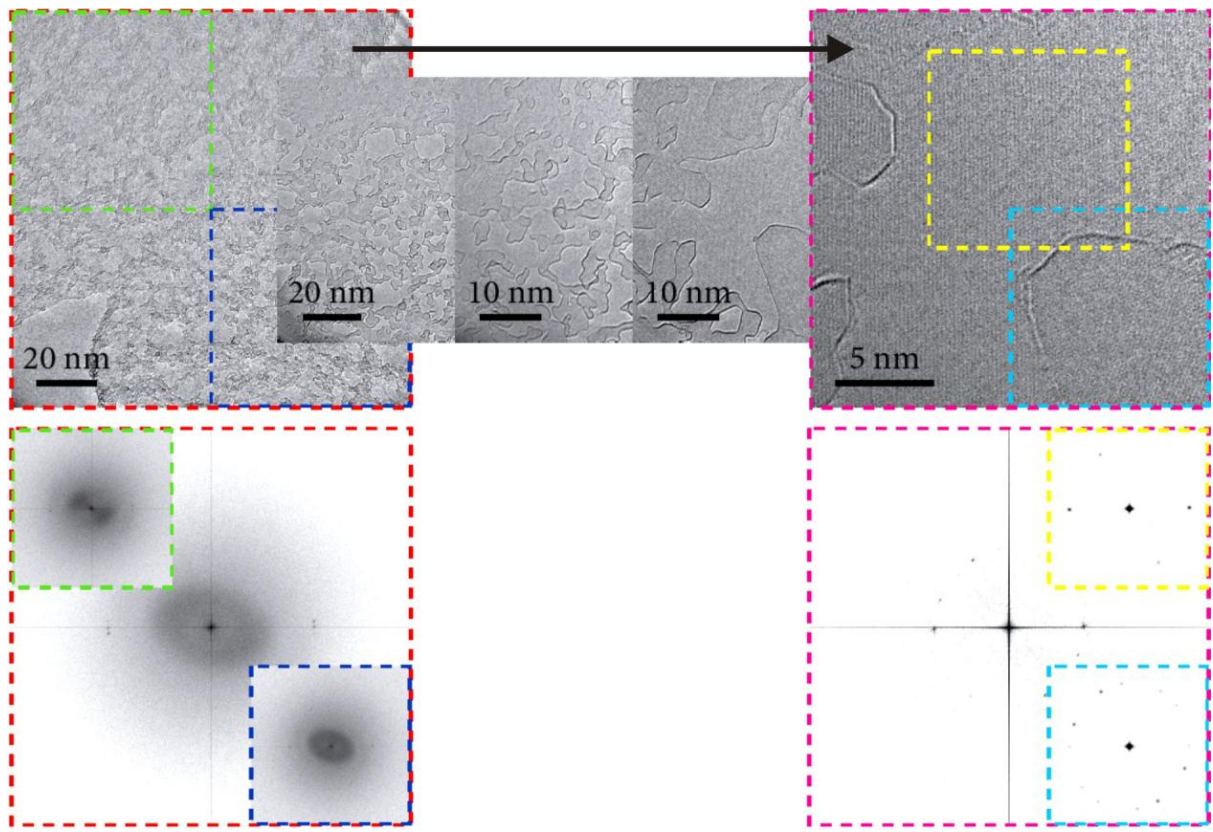

**Figure S3**. Transformation of a-C into graphene. Upper panel. TEM images of the stepwise transformation of a-C into graphene patches by means of current annealing. The black arrow indicates their sequence in time. The first (big) micrograph on the left was formed by two deparated pieces of graphene that fused together [8]. Moreover, both of them are covered with a-C.
Lower panel. Corresponding Fourier Transforms (FT) of the first and last (big) TEM micrographs in the upper panel. (Left) The two separate graphene pieces that fused lead to two FT patterns – one for each side. The big image corresponds to the FT transform of the complete micrograph, while the two smaller ones correspond to the regions marked by the green and blue squares in the separated layers. (Right) The big image corresponds to the FT transform of the complete micrograph, while the two smaller ones correspond to the regions marked by the yellow and orange squares. In the region marked by the orange square we can observe the „original" graphene FT pattern of the lower flake, while in the region marked by the yellow square we can distinguish an additional FT pattern arising from the newly grown graphene patch. The newly grown graphene patch has a different orientation as compared to both of the initial graphene flakes.

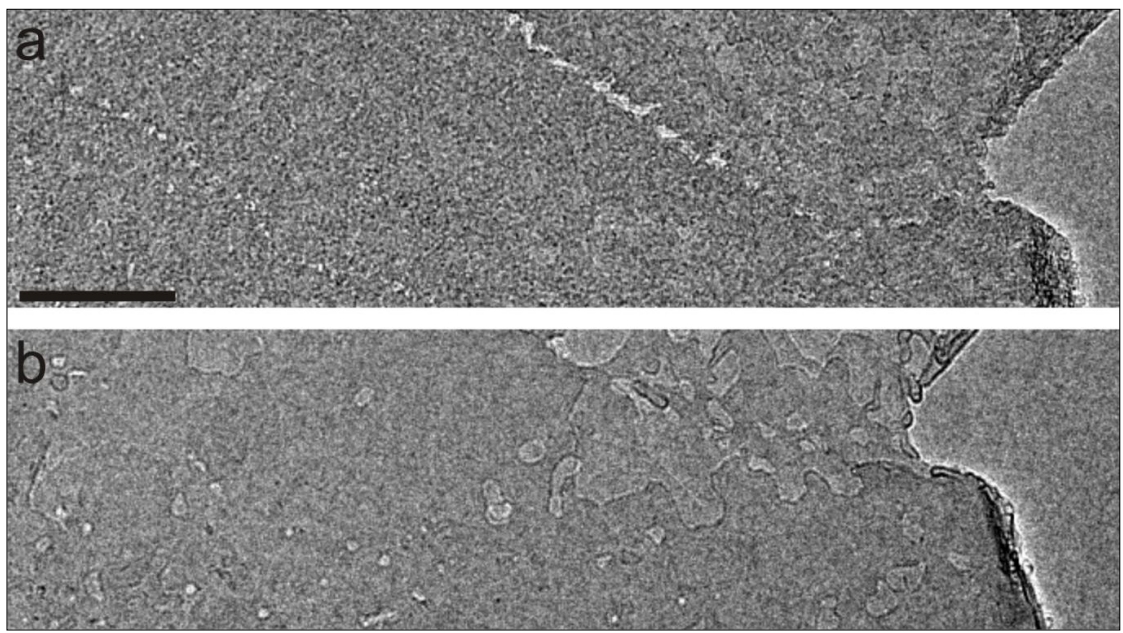

**Figure S4**. Overview of the transformation of a-C into graphene. (a) TEM image of a-C on graphene. The scale bar is 50 nm. (b) TEM image of a layer of graphene grown from a-C that is more than 100 x 300 nm in size and only contains few defects, showing that this process could be scalable.

**Section 5. Healing of holes**

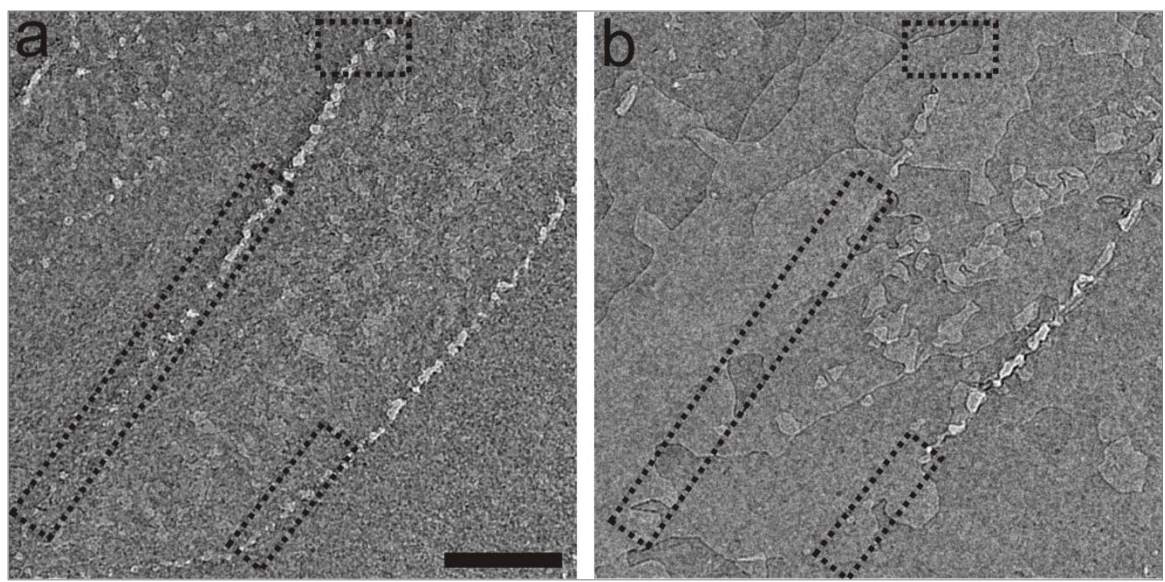

**Figure S5**. TEM images of the gradual healing of holes (bright spots) in the few-layer graphene terraces by means of current annealing in the presence of a-C at 3 V, 0.9 mA. The dotted boxes enclose regions where holes healed out. The scale bar is 50 nm. The time elapsed between (a) and (d) is 3min 30s.

**Section 6. Molecular Dynamics Simulations**

We conducted a series of molecular dynamics (MD) simulations to gain further insight into our experimental findings regarding, first, the transformation of a-C to graphene at high temperatures on top of a graphene substrate and, second, on the healing of holes in a graphene lattice at high temperatures with an a-C feedstock. Born-Oppenheimer G-point molecular dynamics with the non-consistent charge (NCC) - density functional tight-binding (DFTB) scheme [9] were performed. A canonical Nosé-Hoover constrained (NVT) ensemble with a 300 K bath temperature and Nosé-Hoover chains with 100 fs coupling rate were used for initial equilibration (20 ps). The stability of the system was then studied by means of heating steps. Minor deviations between the NCC- and self-consistent charge (SCC) – DFTB [10] results found during the initial equilibration at 300 K justified the use of NCC-DFTB method for all further studies. A standard van der Waals term was included in our DFTB scheme by fitting the London dispersion formula [11]. The system's parameterization for a classical guess was done based on the Tersoff theory [12]. The dynamical trajectories are described by a classical Lagrangian with an isothermal constraint condition given by the Nosé-Hoover theory [13] applied to a system containing N particles, a velocity Verlet integrator, using a time step of 1 fs (100 fs coupling rate and 3 chains). Trajectory analysis was done using the Visual Molecular Dynamics (VMD) interface [14].

6.1 Transformation of a-C to graphene

In the experiments, the graphene that acts as a substrate heats up to very high temperatures due to current-induced annealing. The temperatures achieved were estimated to be around 2000 - 3000 ºC [2, 3, 4]. To reproduce this situation in the theoretical modeling we assume a canonical ensemble as the heat source for the whole system at a constant temperature. We conducted MD simulations employing module DFTB embedded in a CP2k package [15] of a perfect 30x30 $Å^2$ graphene substrate and four a-C clusters of 1 nm diameter on top. These a-C clusters are arranged in tetragonal packing on top of the graphene substrate at a distance of ~ 3.5 Å.

We first turn to the simulation of the clusters of a-C. According to P.W. Anderson [16] a-C is restricted to the description of carbon materials with a localized π-electron system, and in each particular case the $sp^2/sp^3$ ratio should be specified explicitly. In our experiments, the $sp^2/sp^3$ ratio cannot be determined experimentally and we modeled it to be 50/50. Each isolated cluster with randomly chosen atomic coordinates satisfying this ratio was first annealed using the non-bonded potential given by Tersoff theory [12] coupled to 300 K NVT (Nosé-Hoover constrain, 3 chains, 100 fs coupling).

During the MD simulations, the 30x30 $Å^2$ graphene substrate and the four a-C clusters on top are subject to stepwise increasing temperatures. In particular we chose 300, 600, 1200 and 1800 K. The temperatures in the simulations are achieved by a Nosé-Hoover (NH) thermostat [13] with 15 ps annealing at each step, in the frame of the NCC-DFTB model. In the simulations the next temperature step was applied when the potential energy per atom of the a-C cluster subsystem didn't change significantly on the time frame of 2 ps and no more structural changes occurred at that temperature because there is a local minimum of the relative potential energy per carbon atom of the a-C. We used the stepwise annealing strategy to slowly overcome these energy barriers by means of the energy flux provided by the thermostat. The simulations were reproduced 5 times starting with the same geometry but with different initial velocities. A representative example of the stepwise annealing and the corresponding structural changes of the a-C are shown in "Figure S6". Upon increasing temperature, the a-C begins to transform, goes through a glasslike phase and finally forms a graphene lattice. Notably, whereas in the range between 600 and 1200 K the elements produced from the initial a-C clusters remain separated from each other, at 1800 K the elements start to self-assemble and finally produce a single graphene flake. The structure formed is flat and is located 3.5 A above the initial graphene template, similar to the graphite interlayer distance. However, the top crystallographic orientation of the top layer is not aligned with that of the bottom layer, similar to what we observe in the experiments.

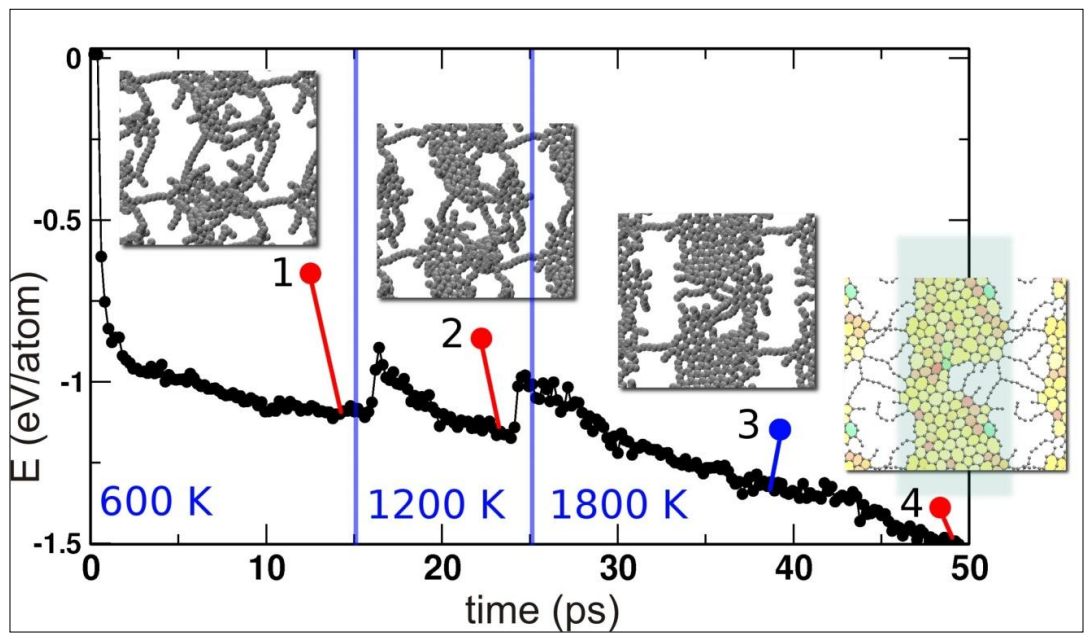

**Figure S6**. MD simulations illustrating the stepwise transformation of a-C to graphene at different temperatures. Potential energy changes of the a-C subsystem for a representable run. Insets 1-4 represent on-time geometries for t = 12.5, 24, 38 and 49 ps, respectively.

When the initially well-equilibrated system at 300 K was directly coupled to a thermostat at 1200 – 1800 K (with 100 fs coupling rate, with integration step 1 fs among the trajectory) in one step, it resulted in an overheating of the a-C and carbon dimer cloud formation leading to an explosion. After such an event, the system required a long time to return back into a compact structure.

However, if the thermal energy is supplied step by step (300, 600, 1200 and finally 1800 K) the system, consisting of the graphene substrate and the 4 a-C clusters on top, remains compact and only internal quick transformations and rearrangements occur. Keeping the system for a long time at 300 K leads to the formation of long fibers, without many changes in a time span of about 10.0 ps. After drastic and quick changes during the first 2 ps the a-C transformation is relaxed to a "glass" form (a local minima) where it stays constant, see "Figure S6 inset 1". This can be understood in terms of the relative energy of the a-C fragment reaching a plateau after several picoseconds of annealing.

Raising the temperature to 1200 K leads initially again to quick rearrangements. The elements distributed over isolated islands are reminiscent of conjugated cage-like structures and are topologically close to a polyaromatic system, "see Figure S6 panel 2". Nevertheless, again the transformation slows down after several ps.

After the next temperature step to 1800 K, the system evolves faster, as can be seen from the increased potential energy slope, "see Figure S-6 (panels 3 and 4)" until graphene is formed which sticks at a distance of 3.5 A to the graphene template. Introducing additional amounts of a-C helps to heal long defects out, as all the a-C initially introduced is not available anymore to be integrated into the forming graphene lattice.

These MD simulations nicely illustrate the role of our graphene template, which prevents the formation of fullerene- like structures.

6.2 Healing of holes in graphene at high temperatures

We simulated a hole of 1 nm radius in an ideal graphene flake and placed 3 a-C clusters (1 nm$^3$) on top of it. In total we followed 5 independent trajectories with different structures and initial velocities. For all these independent MD runs we observed that the hole is healed completely but contains at least one Stone-Wales defect [17], "see figure S7". The formation of Stone-Wales defects is promoted by radicals in a reaction atmosphere [18]. Once these defects are formed, they are extremely stable and an energy of about 4.6 – 7 eV [19, 20] is required to transform the defect to hexagonal sp$^2$ bonds. During the MD runs it will be very rare to reach such high energies during a time span of 30 ps such as in our calculations. Much longer timescales are not feasible from a computational point of view. Nevertheless, it seems that in the real experiment, Stone-Wales defects are healed out during the rather long experiment time.

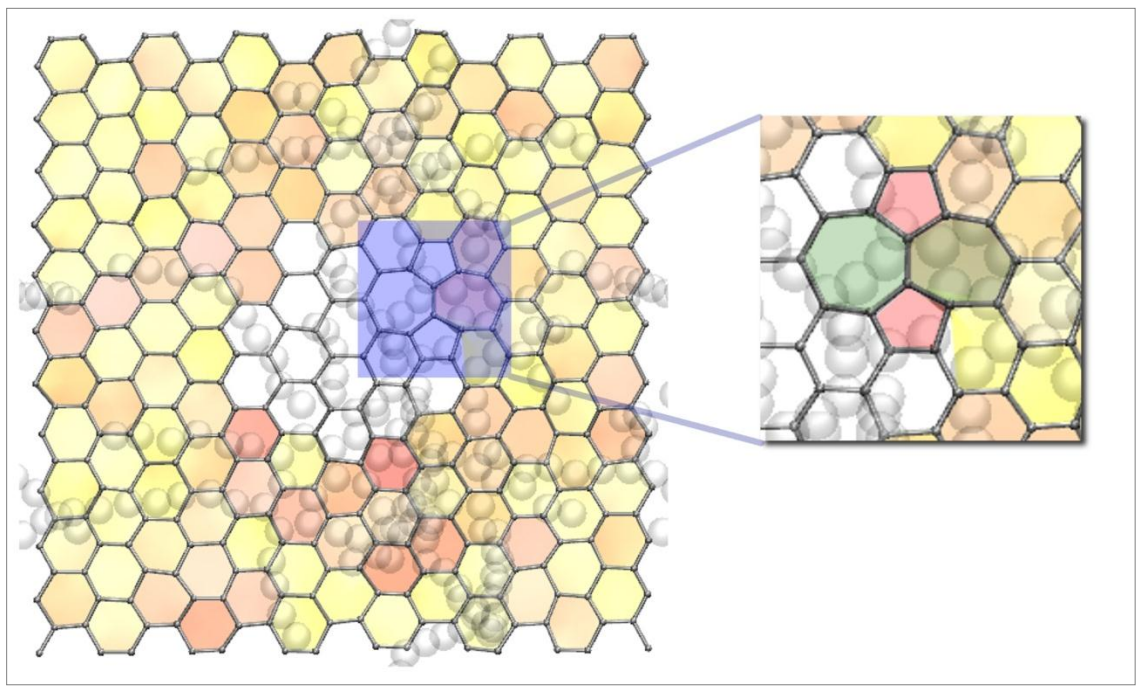

**Figure S7**. Molecular dynamics simulations at 1800 K showing the repaired hole in figure 8. Zoom-in into a Stone-Wales defect. The colors represent the actual number of vertices of the rings and their bending order. The full color code is described in Table 1 on page 134 in Ref. 21.